\documentclass[]{emulateapj}

\slugcomment{Submitted to The Astrophysical Journal}

\shorttitle{SMBH binary evolution having an ultramassive black hole}
\shortauthors{F. M. Khan, Kelly-Holley Bockelmann \& Peter Berczik}

\begin{document}


\title{Ultramassive Black Hole Coalescence }

\author{Fazeel Mahmood Khan\altaffilmark{1}}
\author{Kelly Holley-Bockelmann\altaffilmark{2,3}}

\author{Peter Berczik\altaffilmark{4,5,6}}

\altaffiltext{1}{Department of Space Science, Institute of Space Technology, PO Box 2750 Islamabad, Pakistan;khan@ari.uni-heidelberg.de}
\altaffiltext{2}{Department of Physics and Astronomy, Vanderbilt University, Nashville, TN; k.holley@vanderbilt.edu}
\altaffiltext{3}{Department of Physics, Fisk University, Nashville, TN}
\altaffiltext{4}{National Astronomical Observatories of China,
Chinese Academy of Sciences, 20A Datun Rd., Chaoyang District,
100012, Beijing, China}

\altaffiltext{5}{Astronomisches Rechen-Institut, Zentrum f\"ur
Astronomie, University of Heidelberg, M\"onchhofstrasse 12-14,
69120, Heidelberg, Germany}

\altaffiltext{6}{Main Astronomical Observatory, National
Academy of Sciences of Ukraine, 27 Akademika Zabolotnoho St.,
03680, Kyiv, Ukraine}

\begin{abstract}

Although supermassive black holes (SMBHs) correlate well with their host galaxies, there is an emerging view that outliers exist. Henize 2-10, NGC 4889, and NGC1277 are examples of SMBHs at least an order of magnitude more massive than their host galaxy suggests. The dynamical effects of such ultramassive central black holes is unclear. Here, we perform direct $N$-body simulations of mergers of galactic nuclei where one black hole is ultramassive to study the evolution of the remnant and the black hole dynamics in this extreme regime. We find that the merger remnant is axisymmetric near the center, while near the large SMBH influence radius, the galaxy is triaxial. The SMBH separation shrinks rapidly due to dynamical friction, and quickly forms a binary black hole; if we scale our model to the most massive estimate for the NGC1277 black hole, for example, the timescale for the SMBH separation to shrink from nearly a kpc to less than a pc is roughly 10 Myr. 
By the time the SMBHs form a hard binary, gravitational wave emission dominates, and the black holes coalesce in a mere few Myr.  Curiously, these extremely massive binaries appear to nearly bypass the 3-body scattering evolutionary phase. Our study suggests that in this extreme case, SMBH coalescence is governed by dynamical friction followed nearly directly by gravitational wave emission, resulting in an rapid and efficient SMBH coalescence timescale. We discuss the implications for gravitational wave event rates and 
hypervelocity star production.

\end{abstract}



\section{Introduction}\label{sec-intro}

Almost twenty years after supermassive black holes were discovered in a few nearby galactic nuclei, it is thought that most galaxies host a SMBH with a mass between $10^5-10^{10}\,M_\odot$~\citep{kr95,FF05,gul09}. The growth and evolution of the SMBH is intertwined with the growth and evolution of its galaxy host, and this is borne out through tight correlations between the SMBH mass and fundamental properties of the galaxy such as bulge luminosity~\citep{MM13}, dark matter halo mass~\citep{FF05}, and stellar velocity dispersion~\citep{FM00,geb00}.  

Recent observations probing both the high and low mass end of the galaxy distribution, however, have uncovered examples of embedded SMBHs that are orders of magnitude more massive than the correlations predict. Henize 2-10, a dwarf starburst galaxy with $10^6\,M_\odot$ SMBH \citep{rei11}, is nearly 100 times more massive than it should be based on its dynamical mass. NGC 4486B, a dwarf elliptical, has a $5\times10^{8}\,M_\odot$ SMBH comprising 10\% of the total galaxy mass~\citep{mag98}. NGC1277, a compact lenticular galaxy, may host the most massive SMBH to date, at $1.7\times10^{10}\,M_\odot$ (\citet{van12}, though see \citet{ems13}, which advocates for a more modest mass of $5\times10^9\,M_\odot$). Current efforts are underway to determine if these represent an extreme class of SMBH or if they are statistical outliers.

When the SMBH is such a significant contribution to the mass of the galactic nucleus, it should exact a profound change on the shape, structure, and dynamics of its host. SMBHs are known to alter the shape of a triaxial galaxy \citep{VM98,KHB02}, and can stabilize a galaxy against bar formation \citep{SS04}. 
It is widely accepted that SMBH feedback can quench star formation  (e.g, \citet{pag12,DS05,bun08}), thereby altering the global star formation rate, gas content, and host baryon fraction.
 It is not clear, however, how SMBH coalescence proceeds in this extreme case. Typically, a galaxy merger is expected to be followed eventually by a binary black hole merger deep within the core of the merger. In this framework, SMBH separation is governed on kiloparsec scales by dynamical friction, until the pair joins and forms a hard binary. The separation on the parsec scale is dictated by 3-body scattering of stars within the binary's loss cone. Once the ejected stars have extracted enough energy from the binary orbit to shrink the separation to roughly
milli-parsec scales, gravitational radiation dominates, and the SMBHs coalesce~\citep{BBR80}.  Naively, the orbit of an ultramassive binary SMBH should stall at the parsec scale, since it would need to eject roughly its own mass in stars to reach the gravitational wave stage, and in some cases that could be over half the bulge mass.  Indeed, this {\it final parsec problem} is known to be especially problematic for massive black holes ($>10^{7.5}\,M_{\odot}$ ) and high mass ratios~\citep{MM05}; this may imply that these ultramassive black holes are unlikely to coalesce. 

In this paper, we simulate the dynamics of an ultramassive black hole during a galaxy merger using direct $N$-body simulations.
We find that the merger remnant is quite axisymmetric well within the SMBH influence radius, but it remains triaxial farther out. Contrary to expectations, the black holes reach the gravitational wave regime very efficiently. Surprisingly, the separation between the binary SMBH is mainly driven by dynamical friction; the 3-body scattering phase is very brief, if it exists at all. By the time
the binary is hard, gravitational radiation is already copious, and the black holes coalesce quickly after. We discuss out experimental setup in Section\ref{gal-set},  the results, including post-newtonian effects in Sections \ref{str} \& \ref{results}, and we discuss the implications in Section \ref{sum}.

\section{Simulation Technique}\label{gal-set}

We launch two equal-mass, SMBH-embedded, gas-free, equilibrium, spherical and isotropic galaxy models on a merger orbit to study how ultramassive black holes affect SMBH binary formation and evolution within a realistic live merger remnant.  
Here, each model represents an elliptical galaxy nucleus, one either embedded with an especially massive black hole, or one hosting a SMBH on the $M_\bullet-\sigma$ relation. To ensure that the binary black hole evolution is well-resolved, 
we zoom-in on the late stages of an equal mass galaxy merger;  the simulation starts when the nuclei are only separated by 750 parsec. The nuclei are set on an initial eccentric orbit of 0.76, consistent with typical merger eccentricities~\citep{wet11}. At first pericenter, the SMBH separation is a mere 100 parsecs. 

We represent the stellar density distribution in each nucleus by a~\citet{deh93} profile:

\begin{equation}
\rho(r)=\frac{(3-\gamma)M_\mathrm{gal}}{4\pi}\frac{r_{0}}{r^{\gamma}(r+r_{0})^{4-\gamma}}, \label{denr}\\
\end{equation}

\noindent where $M_\mathrm{gal}$ is the total stellar mass in the galaxy, $r_0$ is its scale radius, and $\gamma$ is logarithmic slope of the central density profile. Increasing $\gamma$ concentrates more stellar mass in the center.  
We construct the model with a central point mass to represent the SMBH and our stellar velocity distribution is chosen from the equilibrium distribution function of this black hole-embedded model \citep{tre94}.

We create three models to represent the primary galaxy, each of which host an extremely massive black hole. Two of our primary models host a SMBH with $0.6\,M_{\rm bulge}$ -- we identify this model as the ultramassive case. The key difference between the two ultramassive models is $\gamma$, which we vary between 1.0 and 1.5.  Our final primary galaxy model contains a central SMBH with $0.2\,M_{\rm bulge}$, which we denote as the overmassive model; this mass is chosen to be consistent with conservative estimates of the SMBH mass in brightest cluster galaxies \citep{pri09}. Two additional models represent the secondary; each model has a $\gamma=1$ slope, but different SMBHs that bracket the range of possible masses on the $M_\bullet$-$M_{\rm bulge}$ relation~\citep{mer01,har04,sct13}. The galaxy parameters are described in Table \ref{TableA}. The total stellar mass of each model, $G$ and $r_0$  are all set to 1.0.

\begin{table}
\caption{Initial Galaxy Models} 
\centering
\begin{tabular}{c c c c c c c c c}
\hline
Galaxy & Role & $N$ & $\gamma$ & ${{M_\bullet}\over{M_{\rm bulge}}}$ & $e_{\rm init}$ & $r_{\rm half}$ & $R_e$ & $r_{\rm infl}$ \\
\hline
UM & Primary & 256k & $1.0$& $0.6$ & 0.76 & 0.71 & 0.53 & 11\\
UM-cusp & Primary &256k & $1.5$& $0.6$ & 0.76& 0.63  & 0.47& 11\\
OV & Primary &256k & $1.0$& $0.2$ & 0.76& 0.71  & 0.53 & 1.7\\
SM & Secondary & 256k & $1.0$& $0.001$ & 0.76 & 0.71 & 0.53 & 0.045\\
LG & Secondary  & 256k & $1.0$& $0.005$ & 0.76 & 0.71  & 0.53 & 0.12\\
\hline
\end{tabular}\label{TableA}
\tablecomments{Column 1: Galaxy. Column 2: Role in the merger. Column 3: Particle number. Column 4: Density cusp. Column 5 Central SMBH mass in units of the bulge mass. Column 6: Initial eccentricity. Column 7: Half mass radius. Column 8: Effective radius. Column 9: SMBH influence radius.} 
\end{table}

\subsection{Physical Scaling}
We use NGC1277 as a reference to scale our primary galaxy models. We set our mass to the bulge mass of NGC1277,   $\sim3\times10^{10}\,M_{\odot}$ \citep{van12}. To obtain the length scaling, we compare the SMBH influence radius in NGC1277 (565 pc) to that of our UM model ( $\sim 11$ model units); this sets one model unit to $\sim50$ pc. Note that since the SMBH is so prominent, the radius of influence encloses 95 percent of the stellar mass. 
Figure \ref{pro2} shows the stellar mass distribution and density in our galaxy models. 
One time step is 0.031 Myr. We integrate at least until the SMBHs form a hard binary; as we shall see, this often marks the transition to the gravitational wave regime. The longest N-body integration time is ($\sim$1570 model units=50 Myrs) for RUN1 and shortest is for RUN3 ($\sim$305 model units=10 Myrs).
The speed of light is 192.15 in our model units.

\begin{table}
\caption{Galaxy Merger Runs} 
\centering
\begin{tabular}{c c c c c c c}
\hline
Run & Primary & Secondary & Measured & Estimated & Estimated \\
          &            &                    &$t_{\rm 1 pc}$ & $t_{\rm df}$ & $t_{\rm coal}$ \\
\hline
RUN1 & UM & SM & 35 & 49 & 80 \\
RUN2 & UM & LG & 8 & 12 &23 \\
RUN3 & UM-cusp & LG & 5 & 4 & 12\\
RUN4 & OV & LG & 3 & 4 & 55\\
\hline
\end{tabular}\label{TableB}
\tablecomments{Column 1: Merger simulation. Column 2,3: Merging galaxy names. Column 4: Time for SMBH separation to reach 1 parsec. Column 5: Estimated dynamical friction timescale, based on \citet{aj11}. Column 6: Estimated SMBBH coalescence time, based on extrapolating the hardening rate and gravitational radiation emission from equations 2-4. All times are quoted in Myr.}
\end{table}

\subsection{Numerical code} \label{num-code}
We carry out $N$-body integrations using $\phi$-GRAPE+GPU, an updated version of $\phi$-GRAPE\footnote{\tt ftp://ftp.ari.uni-heidelberg.de/staff/berczik/phi-GRAPE/} \citep{harfst}, described in section 2.2 of \citet{kh13}. For the current study, we employ a softening of $10^{-4}$ for star-star interactions and 0 for SMBH-SMBH interactions. The simulations were run on 8 nodes (each containing 4 Graphic Processing Units (GPU) cards) on \textit{ACCRE}, a high performance GPU cluster at Vanderbilt University.

\section{Structure and Morphology of the Merger Remnant} \label{str}

\begin{figure}
\centerline{
  \resizebox{0.85\hsize}{!}{\includegraphics[angle=270]{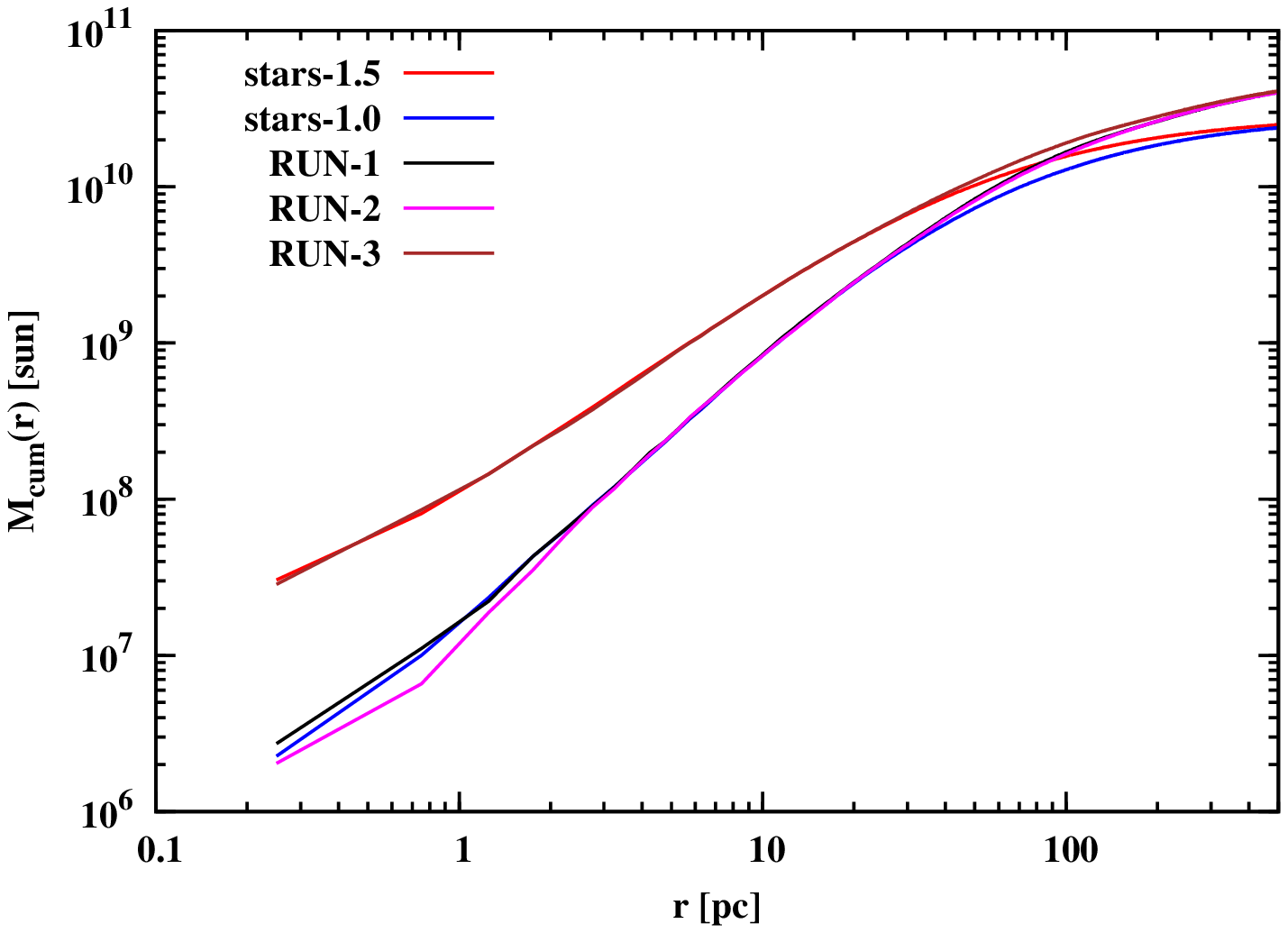}}
  }
\centerline{
  \resizebox{0.85\hsize}{!}{\includegraphics[angle=270]{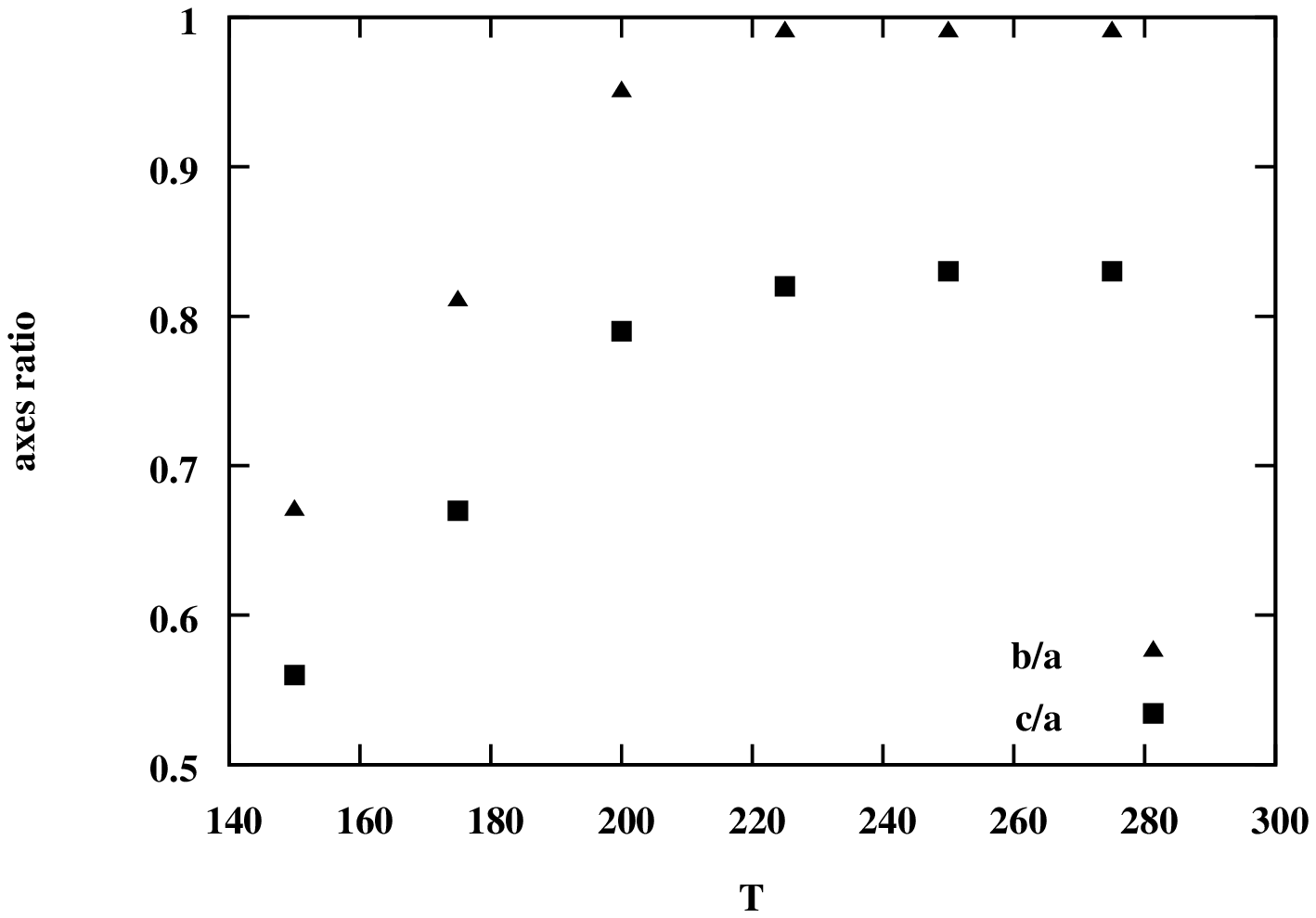}}
  }
\centerline{
  \resizebox{0.85\hsize}{!}{\includegraphics[angle=270]{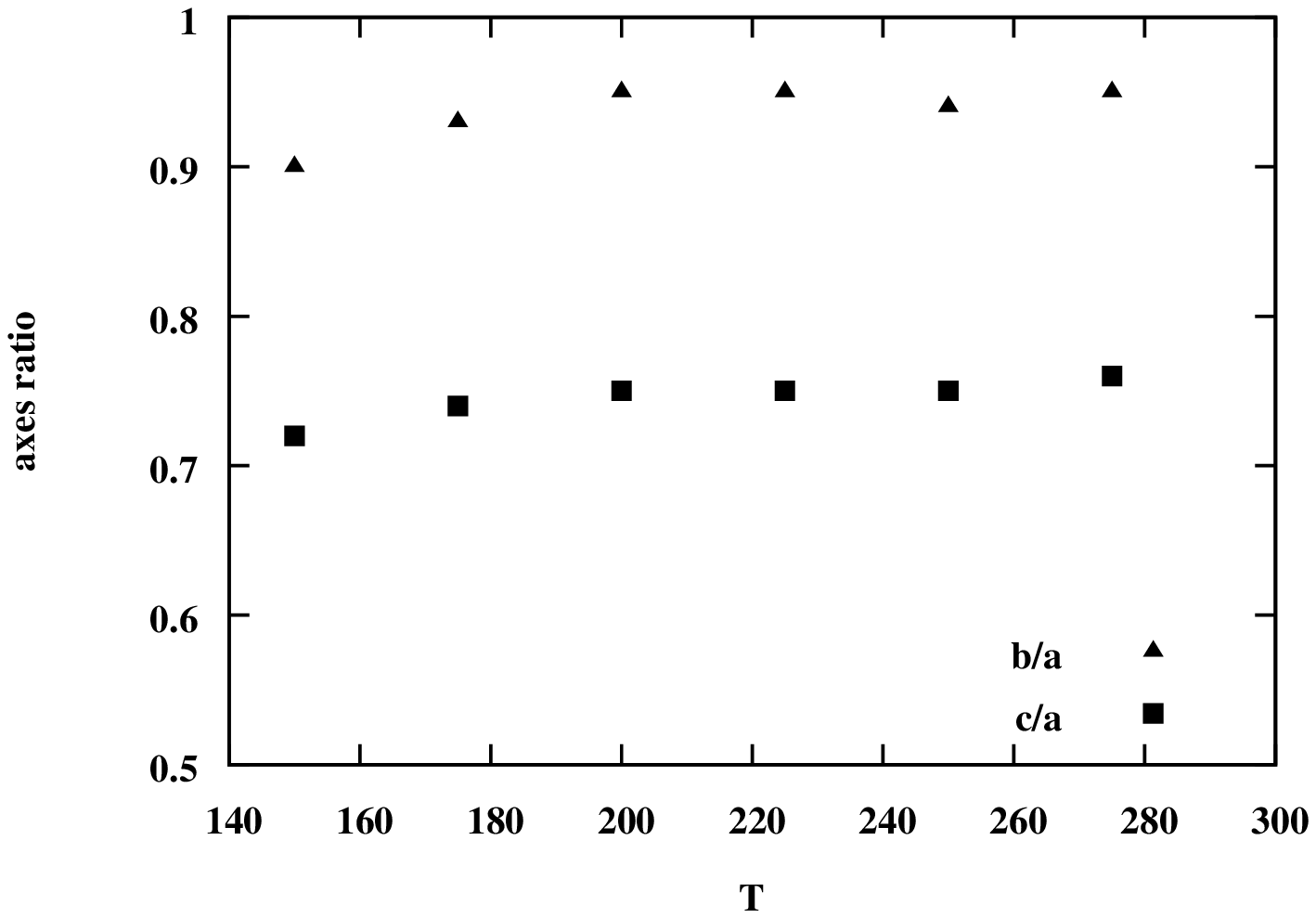}}
  }
  
  \caption[]{
Top: Cumulative stellar mass scaled to NGC 1277. The SMBH dominates the nucleus -- $r_{\rm infl}\sim10\,R_e$ Middle: Evolution of intermediate-to-major (b/a) and minor-to-major (c/a) axes ratios for RUN3 at a radius of 100 pc. Bottom: Same at 500 pc, or $\sim\,r_{\rm infl}$  of the primary SMBH.
} \label{evol}
\end{figure}

Figure~\ref{evol} shows the cumulative mass profile of the merger remnants. RUN1 and RUN2 have similar mass profiles; only the secondary SMBH differs. For RUN3, the steep density cusp is conserved in the remnant, consistent with the interpretation that the three-body scattering phase has not played an active role in sculpting the remnant despite the fact that, as we shall see below, the SMBHs coalesce. With so few 3-body scattering events to scour a core, the most massive SMBH class could reside in cuspy galactic nuclei.

We analyzed the shape evolution of the merger remnant by calculating axes ratios both well inside $r_{\rm infl}$ and at  $\sim\,r_{\rm infl}$. Figure \ref{evol} shows the morphological evolution in RUN3; near the center, the system is initially triaxial, but as it evolves, the triaxiality decreases, and we are left with a moderately flattened axisymmetric central region. However the situation is different at $r_{\rm infl}$ where the system remains stably triaxial. This behavior is consistent with equilibrium studies of SMBH-embedded triaxial galaxies \citep{KHB02,VM98}, where the SMBH induces chaos in the centrophilic orbit population inside $r_{\rm infl}$, a situation that may well be amplified by the 3-body scattering of the SMBH binary.

\section{Dynamics of Ultramassive SMBH Binaries }\label{results}

\begin{figure}
{\centerline{
  \resizebox{0.85\hsize}{!}{\includegraphics[angle=270]{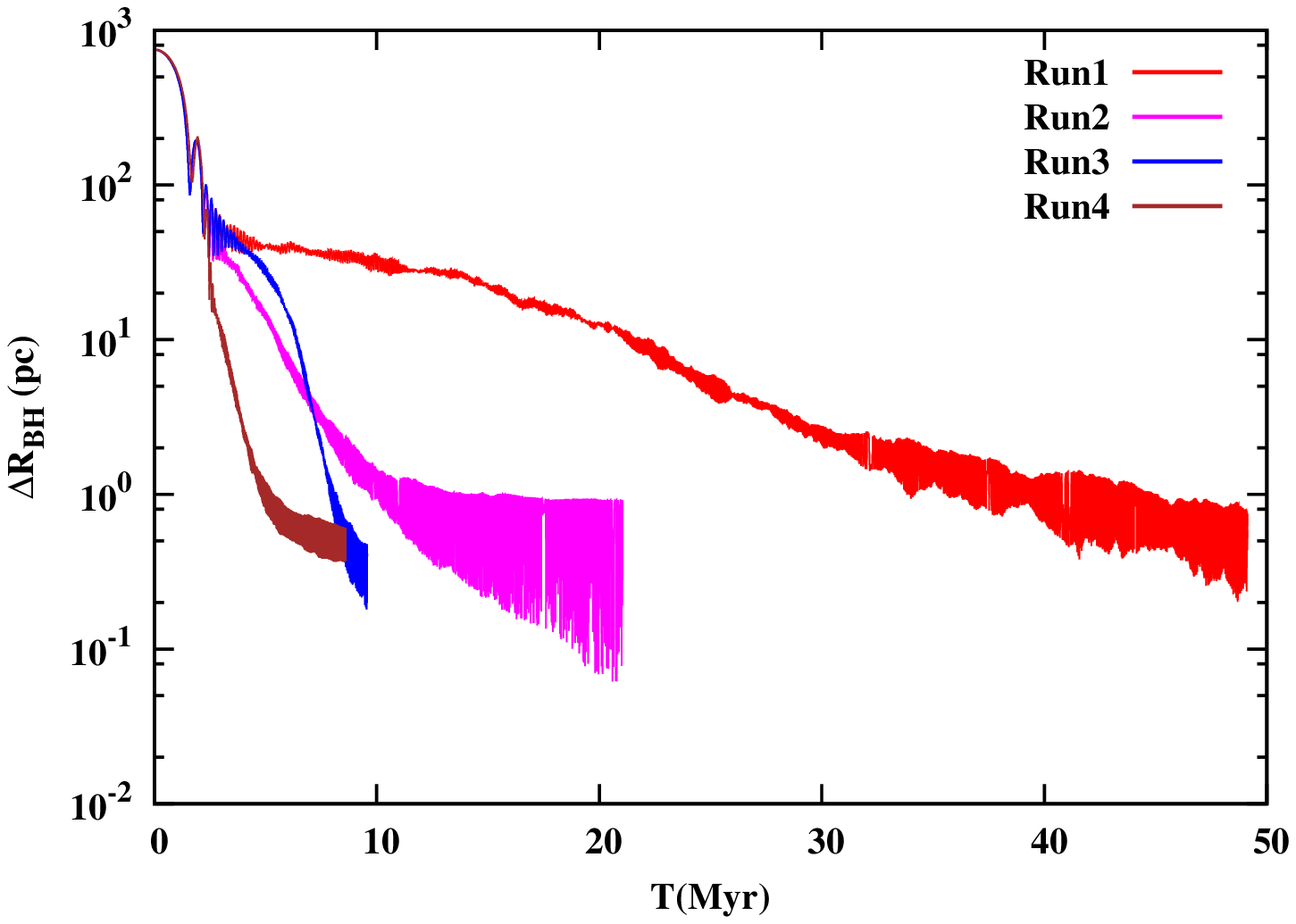}}
  }
\centerline{
  \resizebox{0.85\hsize}{!}{\includegraphics[angle=270]{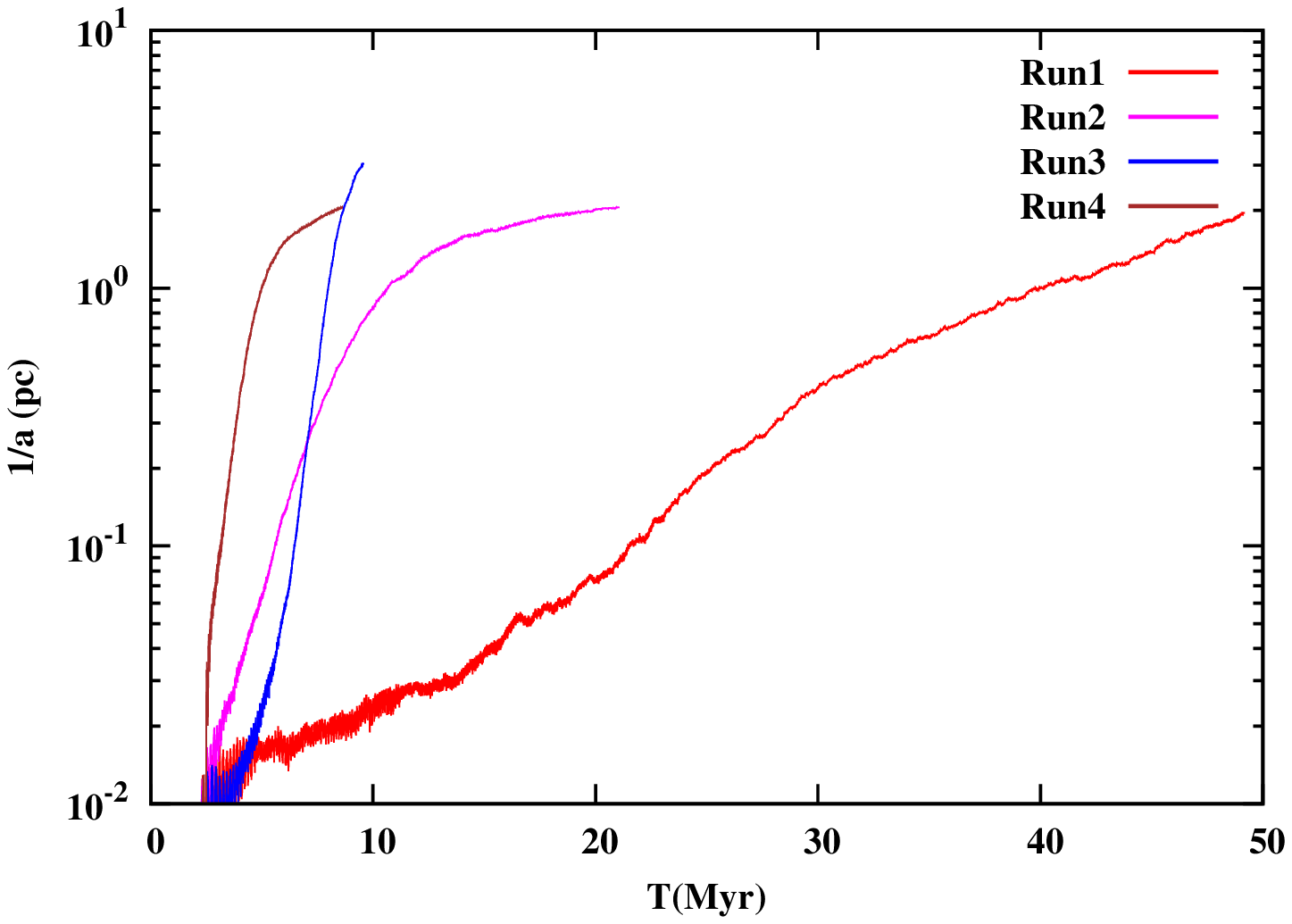}}
  }
  \centerline{
  \resizebox{0.85\hsize}{!}{\includegraphics[angle=270]{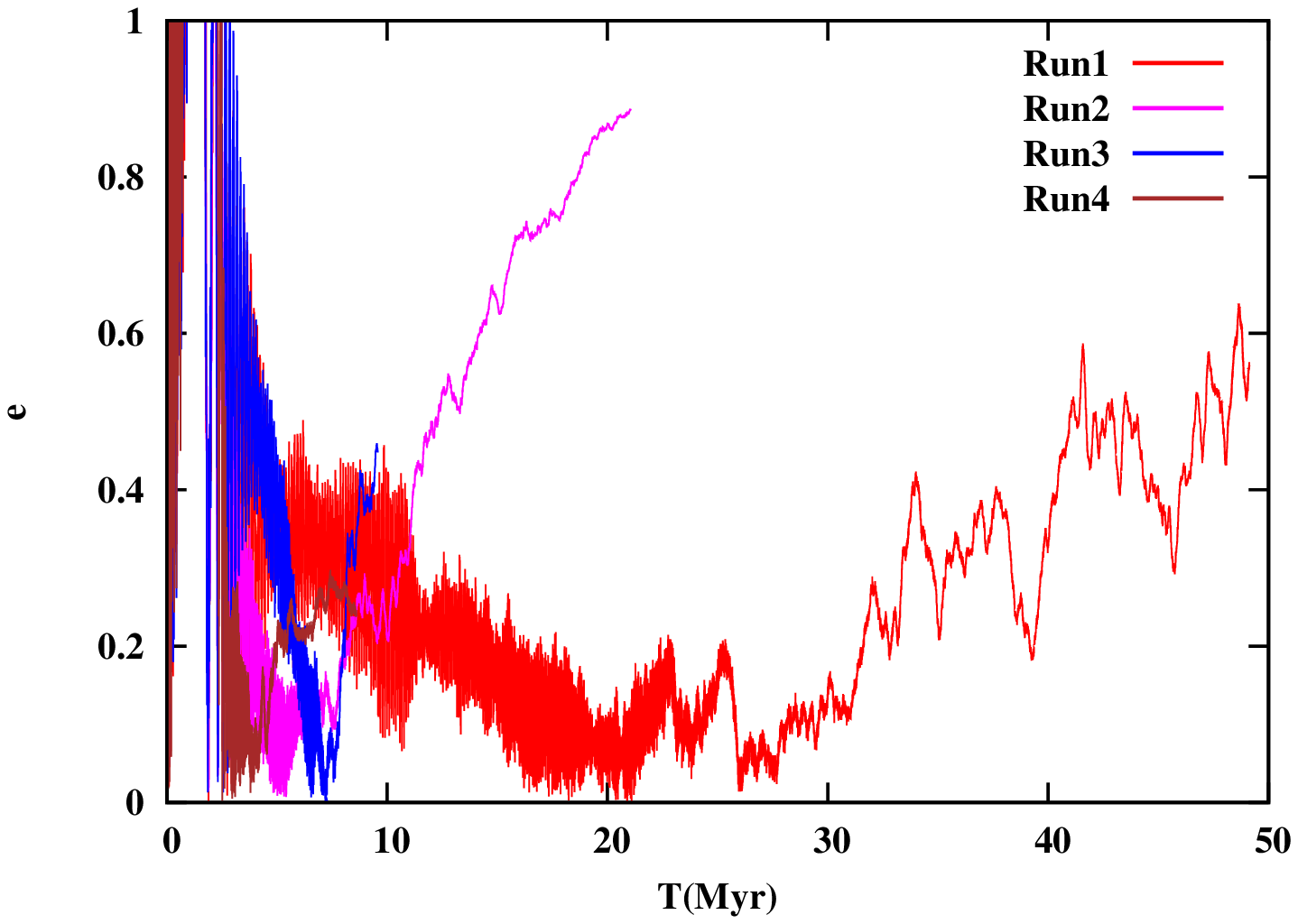}}
  }}
\caption[]{
Evolution of SMBH binary parameters in each galaxy merger simulation: Top: SMBH separation as a function of time. Middle: Evolution of inverse semi-major axes of the SMBH binary, or the hardening timescale. Bottom: Eccentricity evolution of the SMBH binary.}
\label{evol1}
\end{figure}

Here we discuss the formation and evolution of the SMBH binaries in each galaxy merger simulation. Figure~\ref{evol1} shows the binary parameter evolution. We display physical units by scaling the primary to NGC1277 to get a clearer picture of physical domains and time scales involved. To focus on resolving the dynamics of the SMBHs within the remnant, the simulation begins assuming the merger is well underway, and the secondary galactic nucleus has already sunk to the inner kpc of the primary. As a result,  the galaxy merger is complete in only a few Myr; at this time, the density cusp of each galaxy is indistinguishable in phase space.
During this rapid merger phase, the two SMBHs form a binary within the center separated by $\sim\,100$ parsec. 

\subsection{RUN1--UM+SM}

The red line (figure \ref{evol1}, top) shows the evolution of SMBH separation  in RUN1. Here, the primary has a $\gamma=1$ profile and the secondary has a central SMBH mass of 0.001. As expected, dynamical friction governs the early inspiral of the secondary SMBH. 
Initially the inspiral is slow because the background stellar density is low, but as the SMBH moves further in toward the scale radius at 50 pc, the plunge accelerates. The middle panel of this figure shows how the inverse semi-major axis of the binary evolves with time, or the hardening timescale. It is easy to see this initial slow inspiral on the diffuse outskirts of the galactic nucleus. Note that the noise in 1/a is caused by the primary stellar cusp, and reduces as SMBH decouples from the global stellar background and binds to the primary SMBH. 

By 35 Myr, the SMBHs are separated by only 2 pc and the hardening rate slows down, hinting a transition from the dynamical friction phase to three-body scattering phase. The dynamical friction time we observe in the simulation is consistent with the predicted dynamical friction decay time for a SMBH in a SMBH-embedded Hernquist ($\gamma=1$) cusp~\citep{a j11}.  In the dynamical friction phase, the eccentricity (bottom panel of figure) circularizes, reaching an eccentricity $>0.1$, but once the three-body scattering phase begins to dominate, the eccentricity increases very rapidly, reaching $\sim0.6$ at the end of the simulation at 50 Myr.

\subsection{RUN2--UM+LG}

For this run, the secondary SMBH is intended to represent the upper envelope of the $M_\bullet$-$M_{\rm bulge}$ relation and is therefore five times more massive than in RUN1. We expect the SMBH separation to shrink roughly five times faster in the dynamical friction phase, since the background remains same.  Figure \ref{evol1} bears this out; indeed this large SMBH sinks to 1 pc in about 10 Myr (magenta)--approximately five times faster than RUN1. The inverse semi-major axis evolves quickly until the separation is a parsec and then evolves more slowly, indicating a transition from dynamical friction to three-body scattering. Again we see a very rapid increase in eccentricity from below 0.1 to nearly 0.8 when we halt the simulation at 20 Myr because gravitational wave emission dominates the evolution. 

\subsection{RUN3--UMcusp+LG}

Here our primary galaxy mass model is more concentrated, with a $\gamma=1.5$ cusp. This more concentrated model is a better analog to compact ellipticals like NGC 1277~\citep{tri14}. The binary black hole evolution is shown by the blue line in figure \ref{evol1}. The initial secondary SMBH inspiral  is slower than that in RUN2 due to the lower stellar density in the galaxy outskirts. Once the second SMBH enters the primary scale radius ($\sim50$ pc), however, the inspiral rate outpaces RUN2,  and the SMBH separation  shrinks below 1 parsec in less than 10 Myr. This rapid evolution is also reflected in the short hardening timescale of the SMBH binary -- a mere 2 Myr three-body scattering phase ushers the SMBH binary into the gravitational wave regime. Like the previous two runs,  the SMBH binary eccentricity dramatically increases during the 3-body scattering phase.

\subsection{RUN4--OV+LG}

Since there is some debate on the masses of this extreme SMBH class, we use RUN4 to explore a more conservative primary black hole mass; at $0.2\,M_{\rm bulge}$, this SMBH is still a distinct outlier on the black hole-bulge mass relation, but is a factor of 3 smaller than in the ultramassive models. Dynamical friction efficiently shrinks the separation between the two SMBHs down to a parsec. From table \ref{TableB}, we can see that estimated dynamical friction time and observed decay time match very well. Like all previous cases eccentricity increases rapidly at the junction of dynamical friction and three-body scattering phase. The closest comparison is with RUN2; however, the inspiral time is shorter in this case because the satellite suffers less mass loss, thereby behaving as massive `particle' that experiences stronger dynamical friction. The binary forms and hardens at a larger separation. The three-body scattering phase is prolonged compared to RUN2, and the coalescence happens after 50 Myr, roughly twice RUN2.

\subsection{Eccentricity Evolution}

The SMBH binary eccentricity quickly rises in each run, marking when the inspiraling SMBH delves deep enough into core that it encloses a stellar mass roughly equivalent to its own mass. This trend was noticed by \citet{aj11}, and though it marked the transition between dynamical friction and hard binary evolution, the reason was not clear.

\subsection{Modeling the Evolution in the Post-Newtonian Regime}

We can calculate the subsequent SMBH binary evolution with reasonable confidence by following the scheme adopted in \citet{kh12a, kh12b} to model the binary in the Post-Newtonian regime. This, plus the $N$-body evolution allows us to estimate the total SMBH binary coalescence time. We determine the average hardening rate $s=d/dt(1/a)$ during the last few Myr of our simulations, and assume that this rate remains constant until the gravitational wave regime. We also take the final eccentricity $e$ to be constant. SMBH binary evolution can then be estimated as:

\begin{equation}
\frac{da}{dt}=\left(\frac{da}{dt}\right)_\mathrm{NB}+\Big\langle\frac{da}{dt}\Big\rangle_\mathrm{GW}=-s a^{2}(t)+\Big\langle\frac{da}{dt}\Big\rangle_\mathrm{GW}~\label{ratea}
\end{equation}

For gravitational wave hardening, we use the orbit-averaged expression from \citet{pet64}

\begin{mathletters}
\begin{eqnarray}
\Big\langle\frac{da}{dt}\Big\rangle_\mathrm{GW}&=&-\frac{64}{5}\frac{G^{3}M_{\bullet1}M_{\bullet2}(M_{\bullet1}+M_{\bullet2})}{a^{3}c^{5}(1-e^{2})^{7/2}}\times\nonumber \\
&&\left(1+\frac{73}{24}e^{2}+\frac{37}{96}e^{4}\right),~\label{dadt}\\
\Big\langle\frac{de}{dt}\Big\rangle_\mathrm{GW}&=&-\frac{304}{15}e\frac{G^{3}M_{\bullet1}M_{\bullet2}(M_{\bullet1}+M_{\bullet2})}{a^{4}c^{5}(1-e^{2})^{5/2}}
\times\nonumber\\
&&\left(1+\frac{121}{304}e^{2}\right).  \label{dedt}
\end{eqnarray}
\end{mathletters}

\begin{figure}
\centerline{
  \resizebox{0.85\hsize}{!}{\includegraphics[angle=270]{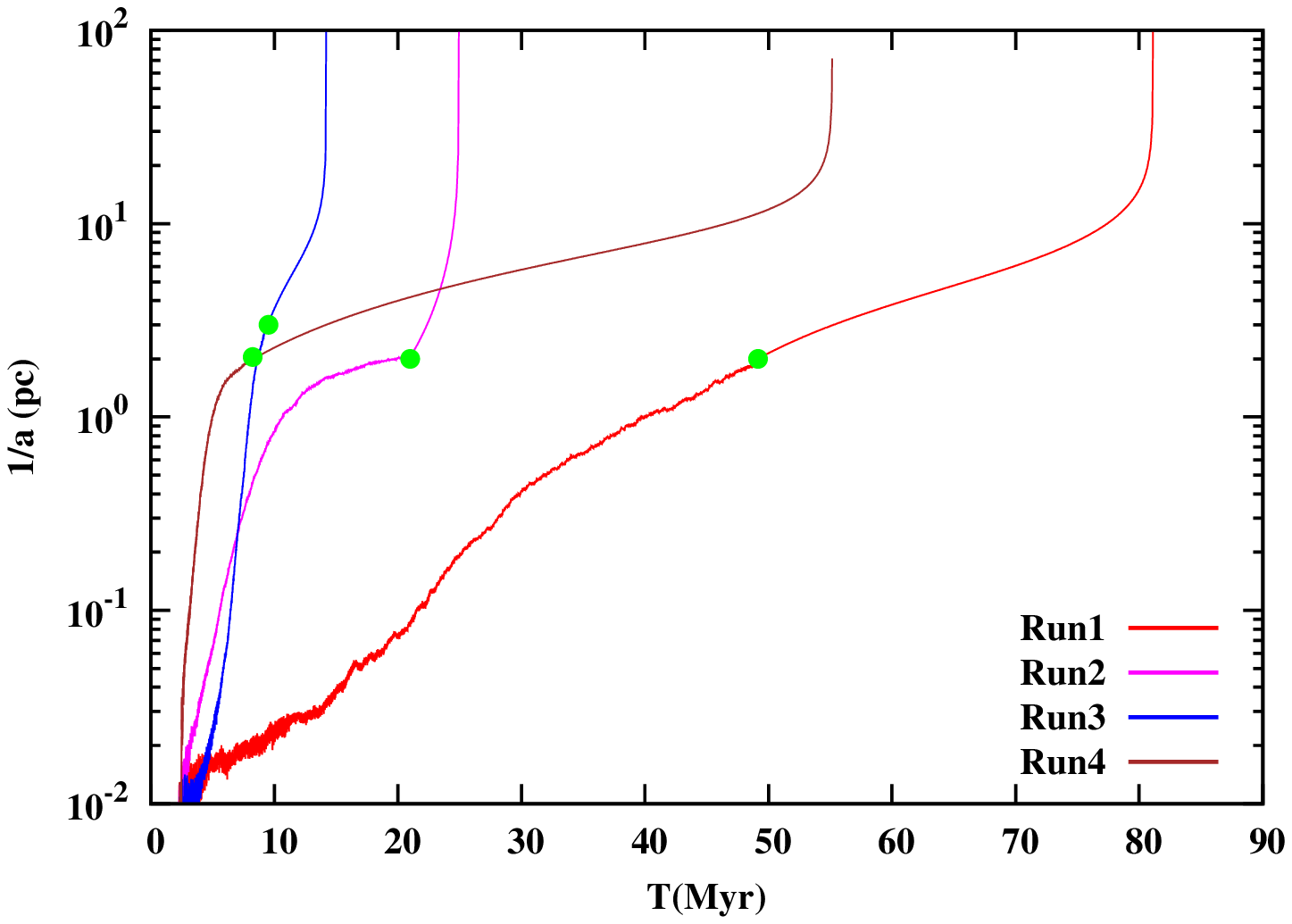}}
  }
  \centerline{
  \resizebox{0.85\hsize}{!}{\includegraphics[angle=270]{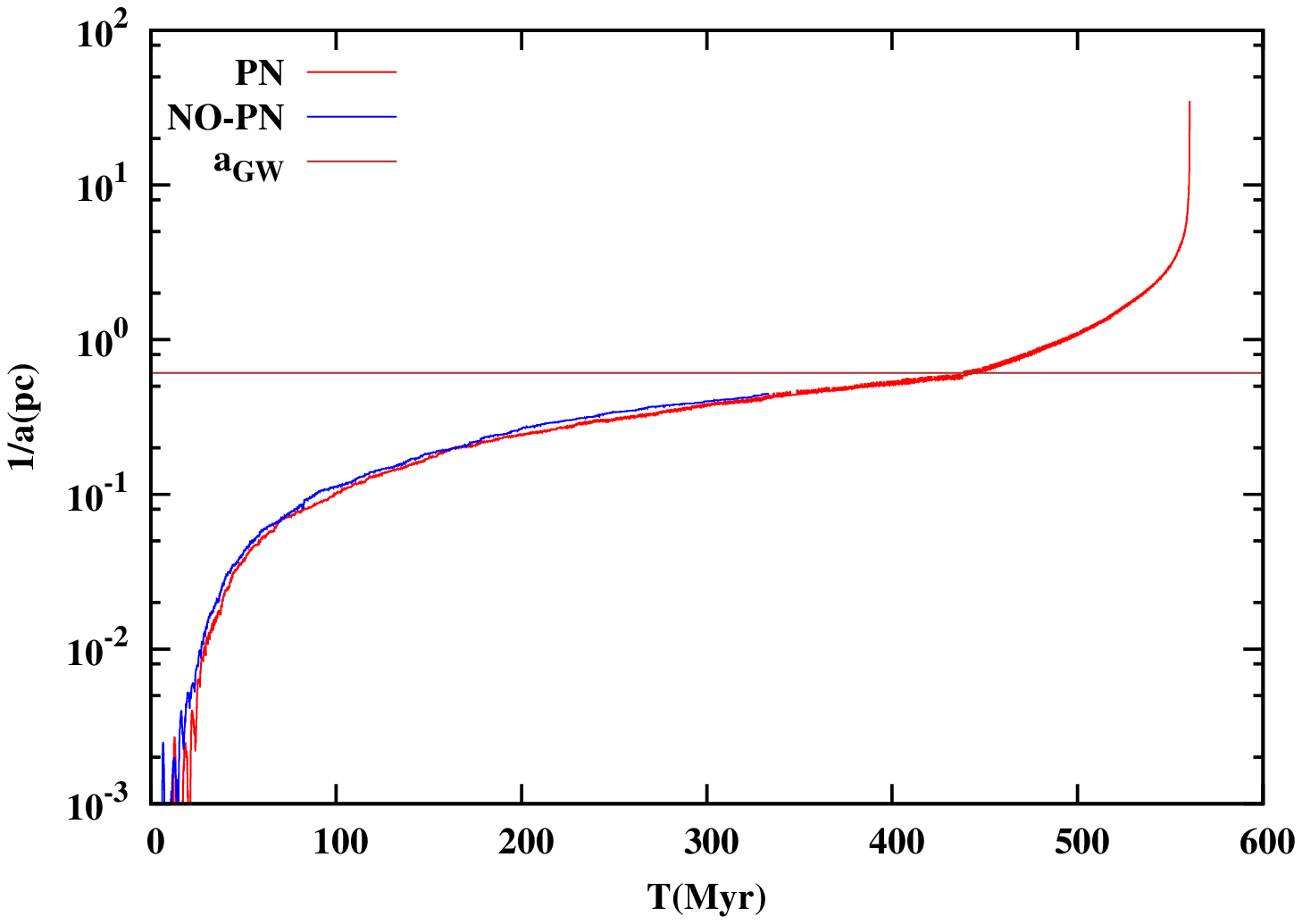}}
  }
\caption[]{
Top: Estimates of SMBH binary hardening in the post-Newtonian regime for all runs. The green dots indicate when the estimate begins.
Bottom: Comparison (A4) from \citep{kh12a}. Blue: SMBH binary evolution from stellar dynamics.  Brown: estimated transition between the stellar dynamical and gravitational wave-dominated regime. Red: SMBH binary evolution including $\mathcal{PN}$ terms.  
} \label{sem2a}
\end{figure}

We solve these coupled equations numerically to calculate the SMBH binary evolution. Estimates of 1/a are shown in figure \ref{sem2a}.  

The SMBH binaries in RUN2 and RUN3 are already in the gravitational wave-dominated regime at the end of our direct $N-$body runs. The total coalescence times, starting from 750 pc when each black hole is embedded in a separate galactic nucleus, are both remarkably small --  23 Myr and 12 Myr, respectively. For RUN4 which has a less massive primary SMBH by a factor of 3, the coalescence is 60 Myr. Three-body scattering phase is prolonged compared to RUN2, and the coalescence takes place after 50 Myr, roughly twice that of RUN2. It appears that when an extreme SMBH is involved in an interaction, the coalescence proceeds  quickly and is mediated predominantly by dynamical friction. For comparison, we include the results of our earlier merger study from \citet{kh12a} where each SMBH is on the $M_\bullet$-$M_{\rm bulge}$ relation (run A4). This run was scaled to M87 with a $3.6\times10^9\,M_{\odot}$ SMBH. We see a long-term 3-body hardening phase, resulting in a coalescence time that is 20-30 times larger than what we witness in this study.

\section{Summary and Conclusion} \label{sum} 

We performed direct $N$-body simulations of the merger of two SMBH-embedded galactic nuclei, in which the primary hosts an extremely massive SMBH. The overall goal was to investigate how these SMBHs affect the structure of the merger remnant, as well as the formation and evolution of the SMBH binary. Though we choose the particular scaling of the primary SMBH to be analogous to the range of estimates for NGC1277, the results are generic for this most massive SMBH class. We followed the late stages of the merger, from kiloparsec separations, through the formation of a bound SMBH binary, to a hard SMBH binary with a separation of less than a parsec. We have two models for the NGC1277 bulge to represent both shallow and cuspier density profiles.

We found that the two nuclei merge in a span of few Myrs and a SMBH binary forms with separation of $\sim100$ pc. Dynamical friction is efficient in driving SMBH binary separation below a parsec. The sinking timescale is more rapid when the secondary SMBH is more massive, or when the primary density profile is higher, since dynamical friction scales with the infalling mass and the stellar background density.  With such large binary separations, this class could represent an excellent prospect for the electromagnetic detection of SMBH binaries, though the lifetime in this phase may preclude direct detection. While at these large separations, gravitational wave emission is already significant; this will likely boost the SMBH binary signal expected by pulsar timing \citep{ra14}.

This dynamical friction phase ends when the inspiralling SMBH sinks deep enough into the center that the enclosed stellar mass is comparable to its own mass. In cases where the secondary SMBH is 0.5 percent of its host bulge mass, dynamical friction is highly efficient, ushering this SMBH so close to the overmassive SMBH that gravitational waves dominate. In these cases, the binary bypasses the 3-body scattering phase seen in typical SMBH binaries.  We expect far fewer hypervelocity stars as a result, and far less significant scouring of the central density profile; indeed, flattened density cusps may not be indicative of this most massive SMBH class. When the inspiralling SMBH is 0.1\% its host mass, there is a brief period of 3-body scattering before the binary enters the gravitational wave-dominated regime. We notice a sudden increase in binary eccentricity at the end of dynamical friction phase.

The merger of two galactic nuclei results in an initially triaxial structure throughout.  Within a dynamical time, the binary SMBH  induces chaos in the centrophilic orbits that define the long axis of the triaxial potential, and with these orbits scattered ergodically, a more axisymmetric figure remains. At SMBH influence radius, the merger product is still triaxial as is expected from previous studies.

Estimated coalescence times of $\sim tens$ Myr are remarkably short compared to all other collisionless studies involving typical SMBH masses. There, the coalescence times are $\sim$0.5-3\,Gyr \citep{kh11,kh12a,kh12b,kh13,gm12,pre11}. Overall, these rapid coalescence times may aid in pinning a future detection of a gravitational wave coalescence to a particular galaxy by requiring that the host is
a recent merger remnant. For this most massive SMBH class, we predict that the black hole merger rate would closely track the host galaxy merger rate, with no significant lag time, and no final parsec problem.

\acknowledgments

 The authors wish to thank the Kavli Institute of Theoretical Physics for hosting an excellent SMBH program, during which much of this paper was hashed out. Simulations were run on Vanderbilt's ACCRE GPU cluster, built and maintained through NSF MRI-0959454, and a Vanderbilt IDEAS grant. KHB acknowledges support from NSF CAREER Award AST-0847696.






\end{document}